\documentclass{article}
\usepackage{graphicx}
\usepackage{amsmath}
\usepackage{harvard}

\input{tcilatex}

\begin{document}

\title{Optimal quantum measurements for spin-$1$ and spin-$\frac{3}{2}$ particles}
\author{P.K.Aravind \\
Physics Department, Worcester Polytechnic Institute, Worcester, MA 01609, USA}
\date{\today}
\maketitle

\begin{abstract}
Positive operator valued measures (POVMs) are presented that allow an
unknown pure state of a spin-1 particle to be determined with optimal
fidelity when $2$ to $5$ copies of that state are available. Optimal POVMs
are also presented for a spin-$\frac{3}{2}$ particle when $2$ or $3$ copies
of the unknown state are available. Although these POVMs are optimal they
are not always minimal, indicating that there is room for improvement.
\end{abstract}

\section{Introduction}

The problem of determining an unknown quantum state of which only a finite
number of copies is available is one that has attracted much attention in
recent years. A fruitful method of attacking the problem was suggested by
Peres and Wootters[1], who pointed out that a judicious joint measurement on
all the copies can sometimes yield more information than any measurements on
the individual copies. This suggestion, whose formal basis is embodied in
the theory of Positive Operator Valued Measurements (POVMs)[2], has served
as the springboard for almost all subsequent work on the problem. Initial
work focussed on determining an unknown (pure) state of a qubit of which $N$
copies were available[3-5], but this work was later extended to mixed states
of qubits[6] and to pure states of qudits (i.e. $D$-state quantum systems)
as well[7,8]. The relationship of quantum state estimation to quantum
cloning machines was discussed in [9]. Other aspects of the state estimation
problem have also been discussed recently[10].

The purpose of this paper is to present some new POVMs that allow optimal
determinations of unknown qudit states in certain cases. Section II presents
POVMs that allow an unknown pure state of a spin-$1$ particle to be
determined with optimal fidelity when $2$ to $5$ copies of that state are
available. Section III presents optimal POVMs for a spin-$\frac{3}{2}$
particle when $2$ or $3$ copies of the unknown (pure) state are available.
The POVMs for spin-$1$ are based on the geometry of the $24$-cell and the $%
600$-cell, two four-dimensional regular polytopes, while the POVMs for spin-$%
\frac{3}{2}$ are based on the $40$ states of the ''Penrose
dodecahedron''[11] as well as another configuration of $60$ states. The
geometric structures underlying the various POVMs presented here have been
exploited earlier to provide proofs of the Bell-Kochen-Specker[12] and Bell
theorems[13,14], and it is interesting that they should prove of use in the
state estimation problem as well.

\section{\protect\bigskip POVMs for spin-$1$}

Given $N$ identical copies of a $D$-state quantum system in an unknown pure
state, how can one determine that state as reliably as possible? We first
address this problem for general $N$ and $D$, introducing our terminology
and notation in the process, before specializing to the case $D=3$ in this
section and $D=4$ in the next. The (now) standard method of attacking this
problem is to implement a POVM (or generalized measurement) on the $N$
copies and to use the outcome of this measurement to make a judicious guess
about the input state. The success of a guess is gauged by a ''fidelity'',
which is generally taken to be the squared modulus of the overlap between
the input state and the guess made for it. The problem of state estimation
consists of devising a POVM for which the average fidel\vspace{0in}ity (i.e.
the fidelity averaged over all possible occurences of the input state) is as
large as possible. If the input state is distributed with uniform
probability over the entire projective Hilbert space of the qudit, it has
been shown[7,8] that the average fidelity is bounded from above by the
quantity $\frac{N+1}{N+D}.$ The task of constructing a POVM that achieves
this upper bound for any $N$ and $D$ was addressed in fairly general terms
in ref.[8], with the analysis for the cases $N=2$ and $3$ (but arbitrary $D$%
) being carried somewhat further. However only for the case $D=3$ and $N=2$
was a POVM explicitly constructed. This section will present optimal POVMs
for $D=3$ and $N=$ $2-$ $5$.

An optimal POVM[2] for arbitrary $N$ and $D$ is characterized by a set of
positive numbers $c_{r}$ and system states $\left| \Psi _{r}\right\rangle $ $%
(r=1,..,k)$ such that

\begin{equation}
\sum_{r=1}^{k}c_{r}\left| \Psi _{r}\right\rangle ^{\otimes N}\text{ }%
^{N\otimes }\left\langle \Psi _{r}\right| =I,
\end{equation}
where $\left| \Psi _{r}\right\rangle ^{\otimes N}$ denotes the tensor
product of the state $\left| \Psi _{r}\right\rangle $ with itself $N$ times
and $I$ is the identity operator in the maximally symmetric subspace of the
space of $N$ qudits. The number of states in the POVM ($=$ $k$) can be
larger than the dimensionality of the maximally symmetric subspace ($=\frac{%
(N+D-1)!}{N!(D-1)!}$) in which the POVM acts. A standard von Neumann
measurement can be regarded as a POVM in which all the $c_{r}$ are unity and
the operators that effect the resolution of the identity are projectors onto
non-overlapping subspaces. A POVM can be implemented in practice by coupling
the system of interest to an auxillary system (the ''ancilla'') and carrying
out a von Neumann measurement on the enlarged system; this measurement then
appears in the space of the system alone as a POVM[2].

The operator identity (1) can be turned into a scalar equation by taking its
expectation value in $|\Psi $ $\rangle ^{\otimes N},$ the $N$-fold tensor
product of the arbitrary system state $|\Psi $ $\rangle $. One then finds
that

\begin{equation}
\sum_{r=1}^{k}c_{r}\text{ }^{N\otimes }\langle \Psi |\Psi _{r}\text{ }%
\rangle ^{\otimes N\text{ \ }N\otimes }\langle \Psi _{r}|\Psi \text{ }%
\rangle ^{\otimes N}=\sum_{r=1}^{k}c_{r}\text{ }\left( \left| \langle \Psi
|\Psi _{r}\text{ }\rangle \right| ^{2}\right) ^{N}=1.
\end{equation}
We now specialize the discussion to a spin-$1$ system for which the spin
value, $J$, is equal to $1$ and the Hilbert space dimension is $D=2J+1=3$.
The projective Hilbert space of a spin-$1$ system has dimension $2D-2=4$ and
an arbitrary state, $|\Psi $ $\rangle $, in this space can be parametrized
as[15]

\begin{equation}
\left| \Psi \right\rangle =\left( e^{i\chi _{1}}\sin \theta \cos \phi
,e^{i\chi _{2}}\sin \theta \sin \phi ,\cos \theta \right) \text{\ \ \ \ \
where \ }0\leq \chi _{1},\chi _{2}\leq 2\pi ,\text{ \ }0\leq \theta ,\phi
\leq \pi /2.
\end{equation}
If one introduces the real variables $x_{1}=\cos \phi \cos \chi
_{1},x_{2}=\cos \phi \sin \chi _{1},x_{3}=\sin \phi \cos \chi _{2}$ and $%
x_{4}=\sin \phi \sin \chi _{2}$ that define a point on the surface of the
four-sphere $x_{1}^{2}+x_{2}^{2}+x_{3}^{2}+x_{4}^{2}=1,$ the state $|\Psi $ $%
\rangle $ can be expressed as

\begin{equation}
\left| \Psi \right\rangle =\left( \sin \theta \left( x_{1}+ix_{2}\right)
,\sin \theta \left( x_{3}+ix_{4}\right) ,\cos \theta \right) ,\ \ 0\leq
\theta \leq \pi /2,\text{ }-1\leq x_{1},x_{2},x_{3},x_{4}\leq 1.
\end{equation}

Our strategy for constructing an optimal POVM for $D=3$ is as follows. For
each $c_{r}$ and $\theta _{r}$ we choose a set of points $%
x_{1i},x_{2i},x_{3i},x_{4i}$ $(i=1,..,n)$ on the surface of the
four-dimensional unit sphere (the same set turns out to suffice for every $r$%
) in such a way that when the sum in (2) is carried out over all $i$ for a
fixed $r$, the dependence on the angular variables $\phi ,\chi _{1}$ and $%
\chi _{2}$ cancels out, leaving a function of $\theta $ alone. The sum on
the left side of (2) then reduces to a polynomial of degree $k-1$ in cos$%
^{2}\theta $ that depends parametrically on the as yet undetermined elements 
$c_{r}$ and $\theta _{r}$ $(r=1,..,k)$ of the POVM. We finally nail down the
POVM by choosing $k$ arbitrary angles $\theta _{r}$ and fixing the constants 
$c_{r}$ in such a way that the polynomial on the left side of (2) reduces
identically to the unity$.$

It remains for us to specify how the points $x_{1i},x_{2i},x_{3i},x_{4i}$ $%
(i=1,..,n)$ should be chosen so that the cancellation over the angular
variables $\phi ,\chi _{1}$ and $\chi _{2}$ can be accomplished on the left
side of (2). For $N=2$ or $3$ it turns out that choosing the $24$ vertices
of a $24$-cell does the trick, while for $N=2-5$ choosing the $120$ vertices
of a $600$-cell does the trick (the $24$-cell and $600$-cell are two of the
four-dimensional regular polytopes[16]). The unit vectors to the vertices of
a $24$-cell are the $16$ vectors $\frac{1}{2}(\pm 1,\pm 1,\pm 1,\pm 1)$ and
the $8$ vectors $(\pm 1,0,0,0),$ $(0,\pm 1,0,0),$ $(0,0,\pm 1,0)$ and $%
(0,0,0,\pm 1)$ (i.e. the vertices of a hypercube plus those of a
hyperoctahedron), while the unit vectors to the vertices of a $600$-cell are
the $24$ vectors just mentioned plus the $96$ vectors obtained by taking all
even permutations of $\frac{1}{2}(\pm \tau ,\pm 1,\pm \tau ^{-1},0),$ where $%
\tau =\frac{1}{2}(1+\sqrt{5})$ is the golden mean.

We now expand upon the above explanation to show how our optimal POVMs for $%
N=2-5$ are constructed$.$ The POVM for $N=2$ is based on three numbers (or
''weights'') $c_{1},c_{2}$ and $c_{3}$ and three angles $\theta _{1},\theta
_{2}$ and $\theta _{3},$ with each angle giving rise to $24$ states based on
the geometry of a $24$-cell as indicated in eqn.(4). On using this POVM in
the left side of (2) and taking $\theta _{1}=\pi /4,$ $\theta _{2}=\pi /2$
and $\theta _{3}=0,$ we find that the summation can be carried out
analytically to yield the polynomial expression

\begin{equation}
2c_{1}+8c_{2}+(8c_{1}-16c_{2})\cos ^{2}\theta +(-4c_{1}+8c_{2}+c_{3})\cos
^{4}\theta .
\end{equation}
For this to be identically equal to $1$, it is necessary that the constant
term be equal to $1$ and that the coefficients of $\cos ^{2}\theta $ \ and $%
\cos ^{4}\theta $ both vanish. This is achieved by the choice of weights $%
c_{1}=\frac{1}{6},c_{2}=\frac{1}{12}$ and $c_{3}=0.$ We have therefore
constructed an optimal POVM for $N=2$ which is characterized by two weights $%
c_{1}=\frac{1}{6}$ and $c_{2}=\frac{1}{12}$ and a total of $48$ states ($24$
arising from each of the angles $\theta _{1}=\pi /4$ and $\theta _{2}=\pi /2$%
). An optimal POVM for $N=3$ can be constructed in a similar manner. This
POVM involves four weights $c_{1},c_{2},c_{3},c_{4}$ and four angles $\theta
_{1},\theta _{2},\theta _{3},$ $\theta _{4}$ and contains $4$ $\times $ $%
24=96$ states obtained by combining each $\theta $ value with the vertices
of a $24$-cell in the manner indicated in (4). On using this POVM with $%
\theta _{1}=\pi /6,\theta _{2}=\pi /4,\theta _{3}=\pi /3$ and $\theta
_{4}=\pi /2$ in the left side of (2) we obtain a third-degree polynomial in $%
\cos ^{2}\theta $ which reduces identically to $1$ if one chooses the
weights as $c_{1}=\frac{2}{27},c_{2}=\frac{1}{18},c_{3}=\frac{2}{9}$ and $%
c_{4}=\frac{7}{108}.$ Our optimal POVM for $N=3$ is therefore characterized
by $4$ weights and $96$ states.

The above construction cannot be used to obtain POVMs for $N\geq 4,$ because
the\ use of a $24$-cell no longer allows the angular dependence on $\phi
,\chi _{1}$ and $\chi _{2}$ to be cancelled out on the left side of (2).
However, as mentioned earlier, replacing the $24$-cell by a $600$-cell
allows this cancellation to be accomplished for all $N$ values from $2$ to $%
5 $. For $N=2$ and $3$ the $600$-cell yields POVMs that are not as
economical as the ones found earlier, so we skip over these cases and pass
to $N=4$ and $5$. The construction of the POVMs for $N=4,5$ proceeds in the
same manner as explained earlier for $N=2$ and $3,$ but with the difference
that the role of the $24$-cell is now taken over by the $600$-cell. Each $%
\theta $ value in the POVM gives rise to $120$ states, and the total number
of states in the POVM is a multiple of $120$. The details of our POVMs for $%
N=4$ and $5 $ (i.e. the weights $c_{i}$ and angles $\theta _{i})$ are
summarized in the bottom half of Table I, while the top half summarizes our
results for $N=2$ and $3.$ For $N=2$ ref.[8] presents a POVM with just $8$
states, which is more economical than the one found here. However for $N\geq
3$ no optimal POVMs of any kind have been reported earlier.

\section{POVMs for spin-$3/2$}

Before presenting our POVMs for spin-$\frac{3}{2}$, we return to eqn.(2) and
recast it in an alternative form based on the Bloch vector description of
qudit systems. In this description[15], the state vector $|\Psi $ $\rangle $
of a $D$-state quantum system is represented by a unit vector -- the
generalized ''Bloch'' vector $\mathbf{\vec{n}}$ \ -- in a real $(D^{2}-1)$%
-dimensional space. The inner product of two state vectors is related to the
scalar product of the corresponding Bloch vectors by the equation

\begin{equation}
\left| \langle \Psi _{i}|\Psi _{j}\text{ }\rangle \right| ^{2}=\frac{1}{2J+1}%
\left( 1+2J\text{ }\mathbf{\vec{n}}_{i}\cdot \text{ }\mathbf{\vec{n}}%
_{j}\right) ,
\end{equation}
where the spin value $J$ is related to $D$ by $D=2J+1.$ On substituting (6)
into (2) and performing some manipulations, (2) can be recast in the form of
a heirarchy of equations that must be satisfied by the POVM elements $%
(c_{r}, $ $\mathbf{\vec{n}}_{r})$ for $r=1,..,k$. We now turn to these
equations[8].

For arbitrary $D=2J+1$ and $N=2$, the equations to be satisfied by the POVM
elements are[8]

\begin{equation}
\sum_{r=1}^{k}c_{r}=(2J+1)(J+1),
\end{equation}

\begin{equation}
\sum_{r=1}^{k}c_{r}\left( \mathbf{\vec{n}}_{r}\cdot \text{ }\mathbf{\vec{n}}%
_{s}\right) =0,
\end{equation}

\begin{equation}
\text{and \ \ }\sum_{r=1}^{k}c_{r}\left( \mathbf{\vec{n}}_{r}\cdot \text{ }%
\mathbf{\vec{n}}_{s}\right) ^{2}=\frac{2J+1}{4J}.
\end{equation}
The two last equations must be satisfied for each value of $s$, so the above
requirements amount to a total of $(2k+1)$ equations.

For arbitrary $D=2J+1$ and $N=3$, the equations to be satisfied by the POVM
elements are[8]

\begin{equation}
\sum_{r=1}^{k}c_{r}=\frac{(2J+3)(2J+1)(J+1)}{3},
\end{equation}

\begin{equation}
\sum_{r=1}^{k}c_{r}\left( \mathbf{\vec{n}}_{r}\cdot \text{ }\mathbf{\vec{n}}%
_{s}\right) =0,
\end{equation}

\begin{equation}
\sum_{r=1}^{k}c_{r}\left( \mathbf{\vec{n}}_{r}\cdot \text{ }\mathbf{\vec{n}}%
_{s}\right) ^{2}=\frac{(2J+3)(2J+1)}{12J},
\end{equation}

\begin{equation}
\text{and \ }\sum_{r=1}^{k}c_{r}\left( \mathbf{\vec{n}}_{r}\cdot \text{ }%
\mathbf{\vec{n}}_{s}\right) ^{3}=\frac{(2J+1)(2J-1)}{12J^{2}}.
\end{equation}
Again the last three equations have to be satisfied for each $s$, leading to
a total of $(3k+1)$ equations altogether. It is interesting to note, from a
comparison of (7)-(9) with (10)-(13), that any POVM for three copies can be
turned into one for two copies simply by reducing each of the constants $%
c_{r}$ by the factor $(2J+3)/3.$ This observation will prove to be of use
below.

We now introduce the 40 states of the ''Penrose dodecahedron''[11] and show
how they can be used to construct POVMs for a spin-$\frac{3}{2}$ particle.
Twenty of these states (termed ''explicit rays'' in [11]) are the spin $+%
\frac{1}{2}$ projections of a spin-$\frac{3}{2}$ particle along the twenty
directions from the center of a regular dodecahedron to its vertices. The
remaining twenty states (termed ''implicit'' rays in [11]) are also
associated with the vertices of the dodecahedron and have the property that
the implicit ray associated with any vertex, together with the explicit rays
associated with the three neighboring vertices, constitute a mutually
orthogonal set of states. Explicit expressions for all 40 Penrose states are
given in Table II, in the basis afforded by four of these states. From this
table it is easily verified that each state is orthogonal to exactly $12$
others and makes the same, constant angle with the $27$ states it is not
orthogonal to. This implies that the scalar products of Bloch vectors for
pairs of Penrose rays have only the two values

\begin{equation}
\mathbf{\vec{n}}_{r}\cdot \text{ }\mathbf{\vec{n}}_{s}=-\frac{1}{3}\text{ \
\ for orthogonal rays,}
\end{equation}

\begin{equation}
\text{or \ \ }\mathbf{\vec{n}}_{r}\cdot \text{ }\mathbf{\vec{n}}_{s}=\frac{1%
}{9}\text{ \ \ \ for non-orthogonal rays. \ \ }
\end{equation}
Using (14) and (15) it is readily verified that a POVM satisfying (10)-(13)
with $J=\frac{3}{2}$ is obtained by taking all forty Penrose rays with a
common weighting factor of $c_{r}=\frac{1}{2}$ for each. Using the remark
just below (13), it follows that a POVM satisfying (7)-(9) is obtained by
taking all forty Penrose rays with a common weighting factor of $c_{r}=\frac{%
1}{4}$ for each. The correctness of these POVMs can also be checked
directly, but much more tediously, by verifying that they make the left side
of (2) reduce identically to the unity for an arbitrary choice of system
state $\left| \Psi \right\rangle .$

We have discovered yet another POVM for $J=\frac{3}{2}$ and $N=2$ or $3$,
consisting of the $60$ states in Table III. For $N=2$ these states are to be
taken with the common weighting factor of $c_{r}=\frac{1}{6}$, while for $%
N=3 $ they are to be taken with the common weighting factor of $c_{r}=\frac{1%
}{3} $. The correctness of these POVMs can be verified in two distinct ways.
The first is to note that each of these $60$ states is orthogonal to $15$
others, makes a constant angle with $32$ others, and a second constant angle
with the remaining$12$ others. In other words, if $\mathbf{\vec{n}}_{1}$%
denotes the Bloch vector of any one of these states and $\mathbf{\vec{n}}%
_{r} $ ranges over the Bloch vectors of all the others, the scalar product
of $\mathbf{\vec{n}}_{1}$ with $\mathbf{\vec{n}}_{r}$ can assume only one of
the following three values:

\begin{equation}
\mathbf{\vec{n}}_{1}\cdot \text{ }\mathbf{\vec{n}}_{r}=-\frac{1}{3}\text{ \
\ \ \ \ for }15\text{ orthogonal rays,}
\end{equation}

\begin{equation}
\mathbf{\vec{n}}_{1}\cdot \text{ }\mathbf{\vec{n}}_{r}=0\text{ \ \ \ \ \ \ \
for }32\text{ non-orthogonal rays, }
\end{equation}

\begin{equation}
\text{and \ \ \ \ }\mathbf{\vec{n}}_{1}\cdot \text{ }\mathbf{\vec{n}}_{r}=%
\frac{1}{3}\text{ \ \ \ \ \ for }12\text{ non-orthogonal rays.}
\end{equation}
Using these scalar products in (7)-(9) and (10)-(13) allows us to confirm
that the $60$ rays, taken with the appropriate weighting factors, are indeed
POVMs for $N=2,3$. The second way of checking the correctness of these POVMs
is to substitute them into the left side of (2) and show that it reduces
identically to the unity for an arbitrary choice of system state $\left|
\Psi \right\rangle $. We say a word about the origin of this POVM: the first 
$24$ states in Table III were introduced by Peres[17] and used by him to
prove the Bell-Kochen-Specker theorem; we added the remaining $36$ states to
obtain a set satisfying all the conditions for a POVM.

This completes our presentation of optimal POVMs for $D=4$ $($or $J=\frac{3}{%
2})$ and $N=2,3$. We have not succeeded in finding any optimal POVMs for $%
D=4 $ and $N\geq 4$.

For $N=2$ the Penrose rays provide an optimal, but perhaps not minimal, POVM
since Acin et.al.[8] constructed a smaller POVM involving only $15$ Bloch
vectors equally inclined to each other; however Acin et.al. did not
demonstrate that physical states corresponding to these Bloch vectors exist,
so their solution cannot be regarded as complete. For $N=3$ the Penrose rays
definitely provide a minimal POVM because it was shown by Acin et.al.[8]
that a minimal POVM in this case cannot consist of fewer than $40$ states.
Our $60$-state POVM, though not minimal, is nevertheless of interest because
of the neat solution it provides to the same problem.

\section{Some open questions}

The optimal POVMs for spin-$1$ presented in this paper all involve a large
number of states. For $N=3$, for example, our POVM involves $96$ states
whereas the minimal number is expected[8] to be a little above $18$. We
suspect, therefore, that there is considerable room for improvement in our
POVMs, as far as their economy is concerned. Realizing this improvement
poses an interesting mathematical challenge, but also one that is not
without physical interest since more economical POVMs would likely lead to
more streamlined experiments.

The POVMs presented in this paper are ''special purpose'' ones, tailored to
specific spin values and small numbers of copies. It is clearly desirable to
generalize the algorithms presented here and devise POVMs for any spin value
and any finite number of copies. In the case of qubits such a general
algorithm was proposed in ref.[4], where it was shown how a POVM for $N$
copies can always be constructed out of a suitable set of $(N+1)^{2}$
states. The technique underlying this construction is to distribute unit
vectors over the Bloch sphere in such a way that a cancellation over the
azimuthal angle $\phi $ is first achieved, following which a cancellation
over the polar angle $\theta $ is achieved by suitably adjusting the
weights, $c_{r}$, in the POVM. A very similar technique of angular
cancellation (first over the ''azimuthal'' angles $\phi ,\chi _{1}$ and $%
\chi _{2}$, followed by a cancellation over the ''polar'' angle $\theta $)
was employed in the construction of our special purpose POVMs for spin-$1.$
We suspect that this technique can be generalized to yield POVMs for any
spin and any number of copies, but the specific way to do this has so far
eluded us.

The problem of realizing the optimal POVMs proposed here as von Neumann
measurements on an enlarged space of the system and an ancilla is an
interesting one worth addressing. This would bring the scheme of optimal
measurements one step closer to experimental realization and also help
highlight any problems connected with its practical implementation.

\textbf{Acknowledgement. }I would like to thank Rolf Tarrach for stimulating
my interest in the problem of optimal quantum measurements and for several
valuable discussions on the subject.

\bigskip \newpage

\bigskip

\begin{tabular}{|l|l|l|l|}
\hline
$N$ & \ \ \ \ \ \ \ \ \ \ \ \ \ \ \ \ \ \ \ Angles $\theta _{r}$ & \ \ \ \ \
\ \ \ \ \ \ \ \ \ \ \ \ \ \ Weights $c_{r}$ & \# states \\ \hline\hline
$2$ \  & $\ \ \ \ \ \ \ \ \ \ \ \ \theta _{1}=\pi /4,\theta _{2}=\pi /2$ & $%
\ \ \ \ \ \ \ \ \ \ \ \ \ \ \ c_{1}=\frac{1}{6},c_{2}=\frac{1}{12}$ & $\ \ \
\ 48$ \\ \hline
$3$ & $\theta _{1}=\pi /6,\theta _{2}=\pi /4,\theta _{3}=\pi /3,\theta
_{4}=\pi /2$ & $\ \ c_{1}=\frac{2}{27},c_{2}=\frac{1}{18},c_{3}=\frac{2}{9}%
,c_{4}=\frac{7}{108}$ & $\ \ \ \ 96$ \\ \hline
$4$ & $\theta _{1}=\pi /6,\theta _{2}=\pi /4,\theta _{3}=\pi /3,\theta
_{4}=\pi /2$ & $\ \ c_{1}=\frac{1}{45},c_{2}=\frac{1}{60},c_{3}=\frac{1}{15}%
,c_{4}=\frac{7}{360}$ & $\ \ \ 480$ \\ \hline
$5$ & $\theta _{1}=\pi /6,\theta _{2}=\pi /4,\theta _{3}=\pi /3,\theta
_{4}=\pi /2,$ & $c_{1}=\frac{2}{225},c_{2}=\frac{17}{300},c_{3}=\frac{2}{75}%
,c_{4}=\frac{29}{1800},$ & $\ \ \ 720$ \\ 
$\ \ $ & $\ \ \ \ \ \ \ \ \ \ \ \ \ \theta _{5}=\pi /8,\theta _{6}=3\pi /8$
& $\ \ \ \ \ \ \ \ \ \ \ c_{5}=\frac{2-\sqrt{2}}{60},c_{6}=\frac{2+\sqrt{2}}{%
60}$ &  \\ \hline
\end{tabular}
\newline
\newline
\\[0.06in]
\ \ \ \ \bigskip

\textbf{TABLE I. Optimal POVMs for estimating an unknown (pure) state of a
spin-}$1$ \textbf{particle of which }$N=2$ $-5$ \textbf{copies are
available. The POVMs for }$N=2,3$ \textbf{are based on the geometry of a }$%
24 $-\textbf{cell, while those for }$N=4,5$\textbf{\ are based on the
geometry of a }$600$-\textbf{cell}. \textbf{The states of the POVMs are
constructed by combining the }$\theta $ \textbf{values in column 2 with the
vertices of a }$24$\textbf{- or }$600$-\textbf{cell in the manner indicated
in eqn.(4), the total number of states in the POVM being either }$24$ 
\textbf{or} $120$ \textbf{times} \textbf{the number of }$\theta $ \textbf{%
values in column 2 (this total is indicated in column 4). The ''weights''
associated with all states sharing a common }$\theta $ \textbf{value are} 
\textbf{indicated in column 3.}

\textbf{\ }

\bigskip \newpage

\newpage

\bigskip

\begin{tabular}{|l|l|l|l|}
\hline
$\left| \Psi _{A}\right\rangle =(1,p,p^{2},0)$ & $\left| \Psi
_{F}\right\rangle =(1,0,0,0)$ & $\left| \Psi _{B}\right\rangle =(0,1,0,0)$ & 
$\ \left| \Psi _{E}\right\rangle =(0,0,1,0)$ \\ \hline
$\left| \Psi _{L}\right\rangle =(-1,0,p^{2},1)$ & $\left| \Psi
_{G}\right\rangle =(0,-1,p,1)$ & $\ \left| \Psi _{C}\right\rangle
=(p^{2},1,0,1)$ & $\ \left| \Psi _{D}\right\rangle =(p,0,1,1)$ \\ \hline
$\left| \Psi _{J}\right\rangle =(0,p^{2},1,-1)$ & $\left| \Psi
_{K}\right\rangle =(1,p^{-2},0,1)$ & $\left| \Psi _{R}\right\rangle
=(0,p,-1,1)$ & $\left| \Psi _{M}\right\rangle =(p^{-1},0,1,1)$ \\ \hline
$\left| \Psi _{H}\right\rangle =(1,0,p,-1)$ & $\left| \Psi _{I}\right\rangle
=(1,p^{2},0,1)$ & $\ \left| \Psi _{P}\right\rangle =(p^{-2},1,0,1)$\ \  & $\
\left| \Psi _{Q}\right\rangle =(0,1,p^{2},-1)$ \\ \hline
$\left| \Psi _{S}\right\rangle =(1,1,0,1)$ & \ $\left| \Psi
_{N}\right\rangle =(0,1,1,-1)$ & $\ \left| \Psi _{U}\right\rangle
=(-1,0,1,1) $\  & $\ \left| \Psi _{T}\right\rangle =(p^{2},p,1,0)$ \\ \hline
$\left| \Psi _{A}^{^{\prime }}\right\rangle =(0,0,0,1)$ & $\left| \Psi
_{F}^{^{\prime }}\right\rangle =(0,p^{2},1,p)$ & $\left| \Psi _{B}^{^{\prime
}}\right\rangle =(p,0,1,p^{2})$ & $\left| \Psi _{E}^{^{\prime
}}\right\rangle =(p^{-2},p^{2},0,1)$ \\ \hline
$\left| \Psi _{L}^{^{\prime }}\right\rangle =(0,1,1,p)$ & $\left| \Psi
_{G}^{^{\prime }}\right\rangle =(1,0,-1,p^{-1})$ & $\left| \Psi
_{C}^{^{\prime }}\right\rangle =(1,0,-1,p)$ & $\left| \Psi _{D}^{^{\prime
}}\right\rangle =(1,1,0,p^{2})$ \\ \hline
$\left| \Psi _{J}^{^{\prime }}\right\rangle =(1,1,0,p^{-2})$ & $\left| \Psi
_{K}^{^{\prime }}\right\rangle =(0,1,1,p^{-1})$ & $\left| \Psi
_{R}^{^{\prime }}\right\rangle =(-1,1,p^{-1},0)$ & $\left| \Psi
_{M}^{^{\prime }}\right\rangle =(-1,1,p,0)$ \\ \hline
$\left| \Psi _{H}^{^{\prime }}\right\rangle =(p^{2},-1,1,0)$ & $\left| \Psi
_{I}^{^{\prime }}\right\rangle =(p,1,-1,0)$ & $\left| \Psi _{P}^{^{\prime
}}\right\rangle =(1,p^{-1},1,0)$ & $\left| \Psi _{Q}^{^{\prime
}}\right\rangle =(1,p,1,0)$ \\ \hline
$\left| \Psi _{S}^{^{\prime }}\right\rangle =(1,p^{2},0,p^{-2})$ & $\left|
\Psi _{N}^{^{\prime }}\right\rangle =(0,1,p^{2},p)$ & $\left| \Psi
_{U}^{^{\prime }}\right\rangle =(p,0,p^{2},1)$ & $\left| \Psi _{T}^{^{\prime
}}\right\rangle =(1,-1,1,0)$ \\ \hline
\end{tabular}

\bigskip

\ \ \ \ 

\bigskip

\textbf{TABLE II. The }$40$ \textbf{states of the ''Penrose dodecahedron''.
The ket vectors corresponding to the states are shown as row (rather than
column) vectors for convenience. The first }$20$\textbf{\ states, without
primes, are the ''explicit rays'' while the last }$20$ \textbf{states, with
primes, are the ''implicit rays''. Each state bears a subscript indicating
the vertex of the dodecahedron with which it is associated (see the papers
in ref.11 for a picture of a dodecahedron with its vertices labelled with
the letters used here). The states }$\left| \Psi _{F}\right\rangle ,\left|
\Psi _{E}\right\rangle ,\left| \Psi _{B}\right\rangle $\textbf{\ and }$%
\left| \Psi _{A^{^{\prime }}}\right\rangle ,$ \textbf{consisting of an
implicit ray and the explicit rays at the three neighboring dodecahedron
vertices,\ form a mutually orthogonal set and are used as a basis in which
the components of all the states are expressed. The symbol }$p$ \textbf{%
stands for} $\exp (i\frac{\pi }{3})$\textbf{\ and a normalization factor of }%
$1/\sqrt{3}$ \textbf{is omitted from many of the states. Note that each
state is orthogonal to exactly }$12$ \textbf{others and makes a constant
angle with the remaining }$27$\textbf{.}

\newpage

\bigskip

\begin{tabular}{|l|l|l|l|}
\hline
$\left| \Psi _{1}\right\rangle =(1,0,0,0)$ & $\left| \Psi _{2}\right\rangle
=(0,1,0,0)$ & $\left| \Psi _{3}\right\rangle =(0,0,1,0)$ & $\left| \Psi
_{4}\right\rangle =(0,0,0,1)$ \\ \hline
$\left| \Psi _{5}\right\rangle =(1,1,1,1)$ & $\left| \Psi _{6}\right\rangle
=(-1,1,-1,1)$ & $\left| \Psi _{7}\right\rangle =(-1,-1,1,1)$ & $\left| \Psi
_{8}\right\rangle =(1,-1,-1,1)$ \\ \hline
$\left| \Psi _{9}\right\rangle =(1,1,1,-1)$ & $\left| \Psi
_{10}\right\rangle =(1,-1,-1,-1)$ & $\left| \Psi _{11}\right\rangle
=(1,-1,1,1)$ & $\left| \Psi _{12}\right\rangle =(1,1,-1,1)$ \\ \hline
$\left| \Psi _{13}\right\rangle =(1,0,1,0)$ & $\left| \Psi
_{14}\right\rangle =(0,1,0,1)$ & $\left| \Psi _{15}\right\rangle =(1,0,-1,0)$
& $\left| \Psi _{16}\right\rangle =(0,1,0,-1)$ \\ \hline
$\left| \Psi _{17}\right\rangle =(1,1,0,0)$ & $\left| \Psi
_{18}\right\rangle =(1,-1,0,0)$ & $\left| \Psi _{19}\right\rangle =(0,0,1,1)$
& $\left| \Psi _{20}\right\rangle =(0,0,1,-1)$ \\ \hline
$\left| \Psi _{21}\right\rangle =(-1,0,0,-1)$ & $\left| \Psi
_{22}\right\rangle =(0,-1,-1,0)$ & $\left| \Psi _{23}\right\rangle
=(-1,0,0,1)$ & $\left| \Psi _{24}\right\rangle =(0,-1,1,0)$ \\ \hline
$\left| \Psi _{25}\right\rangle =(1,i,i,1)$ & $\left| \Psi
_{26}\right\rangle =(1,-i,-i,1)$ & $\left| \Psi _{27}\right\rangle
=(1,-i,i,-1)$ & $\left| \Psi _{28}\right\rangle =(1,i,-i,-1)$ \\ \hline
$\left| \Psi _{29}\right\rangle =(-1,1,-i,-i)$ & $\left| \Psi
_{30}\right\rangle =(-1,-1,i,-i)$ & $\left| \Psi _{31}\right\rangle
=(-1,-1,-i,i)$ & $\left| \Psi _{32}\right\rangle =(-1,1,i,i)$ \\ \hline
$\left| \Psi _{33}\right\rangle =(-1,-i,1,-i)$ & $\left| \Psi
_{34}\right\rangle =(-1,i,-1,-i)$ & $\left| \Psi _{35}\right\rangle
=(-1,-i,-1,i)$ & $\left| \Psi _{36}\right\rangle =(-1,i,1,i)$ \\ \hline
$\left| \Psi _{37}\right\rangle =(1,0,0,i)$ & $\left| \Psi
_{38}\right\rangle =(1,0,0,-i)$ & $\left| \Psi _{39}\right\rangle =(0,1,i,0)$
& $\left| \Psi _{40}\right\rangle =(0,1,-i,0)$ \\ \hline
$\left| \Psi _{41}\right\rangle =(1,0,i,0)$ & $\left| \Psi
_{42}\right\rangle =(1,0,-i,0)$ & $\left| \Psi _{43}\right\rangle =(0,1,0,i)$
& $\left| \Psi _{44}\right\rangle =(0,1,0,-i)$ \\ \hline
$\left| \Psi _{45}\right\rangle =(1,i,i,-1)$ & $\left| \Psi
_{46}\right\rangle =(1,-i,-i,-1)$ & $\left| \Psi _{47}\right\rangle
=(1,i,-i,1)$ & $\left| \Psi _{48}\right\rangle =(1,-i,i,1)$ \\ \hline
$\left| \Psi _{49}\right\rangle =(1,i,0,0)$ & $\left| \Psi
_{50}\right\rangle =(1,-i,0,0)$ & $\left| \Psi _{51}\right\rangle =(0,0,1,i)$
& $\left| \Psi _{52}\right\rangle =(0,0,1,-i)$ \\ \hline
$\left| \Psi _{53}\right\rangle =(1,i,1,i)$ & $\left| \Psi
_{54}\right\rangle =(1,-i,1,-i)$ & $\left| \Psi _{55}\right\rangle
=(1,i,-1,-i)$ & $\left| \Psi _{56}\right\rangle =(1,-i,-1,i)$ \\ \hline
$\left| \Psi _{57}\right\rangle =(1,1,i,i)$ & $\left| \Psi
_{58}\right\rangle =(1,-1,i,-i)$ & $\left| \Psi _{59}\right\rangle
=(1,1,-i,-i)$ & $\left| \Psi _{60}\right\rangle =(1,-1,-i,i)$ \\ \hline
\end{tabular}

\bigskip

\textbf{Table III. A set of }$60$ \textbf{states} \textbf{yielding optimal
POVMs for }$J=\frac{3}{2}$ and $N=2,3$ \textbf{(note that most of these
states are unnormalized). The POVM for }$N=2$ \textbf{is obtained by taking
these states with an equal weighting factor of }$c_{r}=\frac{1}{6}$, \textbf{%
while the} \textbf{POVM for }$N=3$ \textbf{is obtained by taking these
states with an equal weighting factor of }$c_{r}=\frac{1}{3}.$

\end{document}